\documentclass[12pt]{iopart}
\input epsf
\usepackage[dvips]{graphicx}
\usepackage[dvips]{epsfig}
\usepackage{subfigure}

\begin{document}

\title{Multiple micro-optical atom traps with a spherically aberrated laser beam}

\author{P Ahmadi\footnote[1]{To whom correspondence should be addressed
(peyman@okstate.edu)}, V Ramareddy and G S Summy}

\address{Department of Physics, Oklahoma State University,
Stillwater, Oklahoma 74078-3072 }

\begin{abstract}
We report on the loading of atoms contained in a magneto-optic trap
into multiple optical traps formed within the focused beam of a $\rm
CO_{2}$ laser. We show that under certain circumstances it is
possible to create a linear array of dipole traps with well
separated maxima. This is achieved by focusing the
laser beam through lenses uncorrected for spherical
aberration. We demonstrate that the separation between the
micro-traps can be varied, a property which may be useful in experiments which require the
creation of entanglement between atoms in different micro-traps.
We suggest other experiments where an array  of these traps
could be useful.

\end{abstract}


\submitto{\NJP}

\maketitle

\section{Introduction}

The trapping of atoms at the intensity maximum of an optical field
that is far-detuned to the red of an atomic transition has been
the subject of study for almost a decade now
\cite{ChuFORT,Hansch&Weitz_FORT}. There has been a rapid growth of
interest in these far-off resonant optical traps (FORTs) because
of their versatility and wide range of possible applications. For example, they
have been used to create an all optical Bose-Einstein condensate
\cite{ChapmanBEC, CsBEC,YbBEC}, a degenerate Fermi gas
\cite{Thomas_FermiGas}  and an all optical atomic laser
\cite{WeitzPRL_BEC}. These traps in the form of optical lattices
have been employed in theoretical models to open new frontiers in
quantum information research. The proposal of Brennen {\it et al.}
\cite{Brennen} for quantum logic gates using neutral atoms in optical lattices,
provided a way around the decoherence problem which affects
schemes involving  charged particles. They
showed that entanglement between a collection of trapped neutral
atoms can be created with a laser using the induced electric
dipole-dipole interaction. The main difficulty associated with their scheme has been the need to construct
a lattice FORT with sufficient separation between
unit cells to address them individually and with a sufficient volume
to load many atoms at each trapping site. These challenges have been the focus of another series of
experimental efforts. For example, using a holographic phase plate, Boiron {\it
et al.} \cite{Borion} constructed an optical lattice with a period
of 29$\mu \rm m$ using a YAG laser. In other experiments, the
Hannover group have developed a technique using arrays of
microlenses to focus a red detuned laser beam and create a series
of micro-traps for use as quantum memories
\cite{Dumke,Birkl,Buchkremer}. Peil {\it et al.} \cite{peil}
employed two independent optical lattices, whose spatial periods
differ by a factor of three, to load a Bose-Einstein condensate of
$\rm Rb \ 87$ atoms in sites having a separation approximately 30$\mu \rm
m$.

In most FORT experiments atoms are trapped at the intensity
maxima formed by a focused laser beam in either a travelling or standing
wave configuration. In this paper, we
demonstrate a new approach in which the peaks in the diffraction
pattern associated with spherical aberration in the vicinity of
the focal plane of a lens are used to create a linear
array of micro-traps. The primary spherical aberration pattern
close to the focal plane has been studied and well
documented by several authors.
For example, Evans and Morgan \cite{Evance,Evance-prl} theoretically produced the
aberration pattern of a lens that was not corrected for spherical
aberration in order to explain laser induced breakdown in gases,
while Smith \cite{smith} experimentally verified the primary
spherical aberration intensity distribution produced by a lens
uncorrected for spherical aberration. The
spherical aberration in
our experiments is induced by the lenses in the path of a $\rm CO_2$ laser beam.
It will be seen that most of the contribution to the spherical aberration comes
from the final lens (primary lens)
which is employed to focus the $ \rm CO_2$ laser beam onto a cold
atomic cloud. We will show that varying the incident beam size on the
primary lens enables us to change the aberration pattern and hence
control the separation of the micro-traps over a range of about a millimeter.
One of the advantages of our set up is the use of a $\rm
CO_2$ laser as a far off-resonant light source. This considerably
improves the coherence time compared to some of the atom optical
experiments mentioned above which use YAG lasers to create dipole
traps with micron size separation. Furthermore, the ability to vary
the spacing between the micro-traps over a range of about a millimeter
makes the addressing of the individual traps feasible using
the techniques developed by N\"{a}gerl {\it et al.} \cite{Nagerl}.

This paper is constructed as follows. In section 2 we discuss the
theory of the multiple trap potential that is used in simulations.
Section 3 is devoted to the description of the experimental setup. In section 4
we present the experimental data and compare them with our
simulation results. Our suggestions and future plans for the use
of these potentials appear in the conclusion section.

\section{Aberration effect of a lens on the incident laser light near the focus}
Since the potential well depth of any FORT is proportional to the
intensity, we now proceed to calculate the intensity distribution
produced near the focus of the lens used in our experiment.
Following Born and Wolf \cite{born} and Yoshida and Asakura
\cite{yashida}, the intensity close to the focus of a lens for an
incident Gaussian beam is given by,
\begin{equation}
I\left(u,v\right) = {1 \over w^2} \ {\Big |} \int_0^1 \rho \ \rmd
\rho \ \rme^{- \rho ^2 \over \left(w/a\right)^2}
\rme^{-\rmi\left({u \rho ^2 \over 2}+k \beta \rho^4\right)}
J_0\left(v \rho\right){\Big |}^2,
\end{equation}
where $w$ is the spot size on the lens and $\rho$ is the
radial coordinate on the lens normalized to the radius of the lens,
$a$. $v$
and $u$ are the scaled cylindrical radial and axial coordinates of
the image space (with the origin for $u$ at the Gaussian focus)
and are given by
$v= {2 \pi \over \lambda} {a\over R}\sqrt{x^2+y^2}$ and
$u= {2 \pi \over \lambda} \left({a\over R}\right)^2z$.
$k$ is the vacuum wave number, given by $k=2 \pi / \lambda$, where
$\lambda$ is the wavelength of the light used. $R$ is the radius
of the Gaussian reference sphere from the lens, $(x,y)$ are the
cartesian coordinates in the Gaussian image plane and $\beta$ is
the primary spherical aberration coefficient, usually expressed in
terms of the number of wavelengths. This coefficient is additive
over all the elements used in an optical system. Our set up has
three lenses in the path of the laser light (see Fig.\thinspace
\ref{exp_schematic}). The first two lenses constitute a telescope
and the third one (which is  placed inside the
vacuum chamber) we refer to as the primary lens.

Using the thin lens approximation,  we calculate the primary
spherical aberration produced by such a lens of focal length $f$
to be \cite{welfold},
\begin{equation}
\beta ={w^4 \over 32 f^3} \left[\left({n \over n-1}\right)^2 +
{\left(n+2\right) \over n\left(n-1\right)^2}\left(B +
{2\left(n^2-1\right) \over n+2} C\right)^2 - {n \over
n+2}C^2\right],
\label{beta}
\end{equation}
where $n$ is the refractive index of the lens medium and $B$ is
the shape variable given by
$ B=(c_1 + c_2)/( c_1 - c_2)  $
and $c_i = 1/r_i$; $i=1,2$; $r_i$ are the radii of curvature of
the lens surfaces. $C$ is known as the conjugate variable and is
defined as
$ C=(u_1 + u_2)/( u_1 - u_2)$,
where $u_1$ and $u_2$ are the divergence angles of the gaussian
beam before and after passing through the lens. These angles are
given by
$u_i=\lambda/ \pi w_{0i}$; $i=1,2$,
where $w_{01}$ and $w_{02}$ are the minimum beam waists of the beam before and after
the lens. It should be noted that according to the usual sign
convention, if the lens produces a converging beam, then $u_2$ is
negative so that the denominator in the definition of $C$ is not zero in our experiment.

In the experimental situation we wish to model, the separation
of the telescope lenses is varied by moving the second lens in the
optical system and keeping the other lenses fixed. Thus the first
lens of the telescope contributes a constant
amount to the total spherical aberration coefficient $\beta$
as its $w$ and $C$  parameters are fixed.
As the position of the second lens in the telescope is moved, the beam
size on this lens and on the third (the primary) lens will change
resulting in changes to the parameters $w$ and $C$ for these lenses.
This leads to a variable contribution to $\beta$ by these last two
lenses and therefore a variable aberration pattern near the focus of the primary lens.
We have found that for our experiment the primary spherical
aberration of the primary lens varies from around 0.1 up to around 18.2
wavelengths.

\section{Experimental set up}

In the following, we present an experimental  setup which enables us to create  the spherical
aberration pattern to form a series of micro optical traps. Our
experimental apparatus consists of a stainless steel, octagonal
vacuum chamber which is maintained at a pressure of approximately
$\rm 5 \times 10^{-10}$ torr by an ion pump. A 3.1 cm diameter
ZnSe viewport allows us to focus the 10.6 $\mu$m light from a $\rm
CO_{2}$ laser into the center of this chamber. The focusing lens (the primary lens)
is a meniscus lens with a 3.81 cm focal length
and 2.54 cm diameter placed inside the vacuum chamber and is not
corrected for spherical aberration\footnote[1]{Note that for some
of the later experiments, this lens was replaced by an aspheric lens}.
This lens is mounted such that
the convex side is towards the center of the chamber to maximize
the spherical aberration effects. Before reaching the primary lens, the $\rm CO_2$ laser beam passes
through a telescope composed of two plano convex lenses with
6.35 cm and 12.7 cm focal lengths placed approximately 2 meters
away from the chamber. This configuration allows us to control the
beam size at the lens inside the chamber by varying the
separation of the telescope lenses. Consequently, we are able to change the spherical
aberration pattern close to the gaussian focus inside the chamber.

\begin{figure}
\begin{center} \mbox{\epsfxsize 3.0in \epsfbox{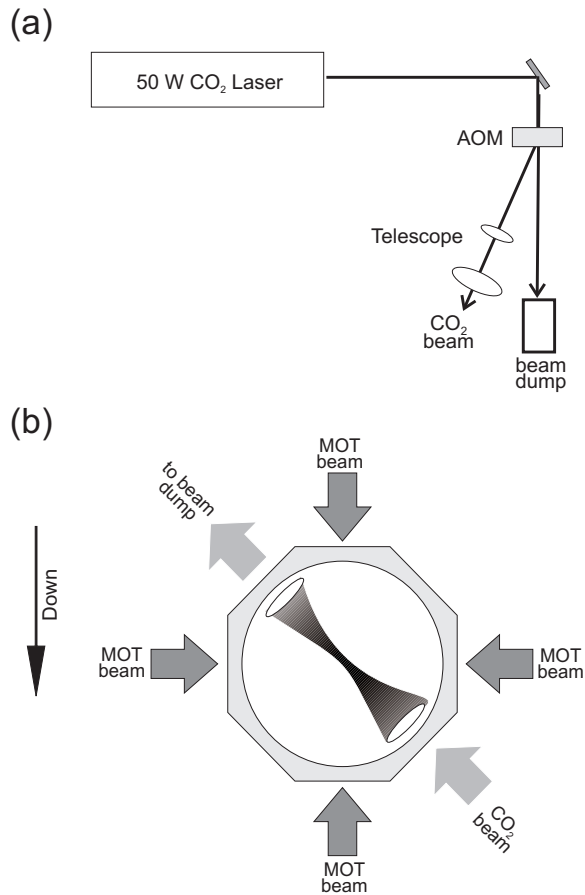}}
\end{center}
\caption{Experiment schematic, showing (a) the production of the FORT beams
and its path before it enters the chamber, and (b) the beam geometry relative
to the vacuum chamber.} \label{exp_schematic}
\end{figure}

The trapping light was directed into the vacuum system in a geometry such that it
propagated at an angle of 45 degrees to the vertical. The light
for this beam originated from a 50 Watt, RF excited $\rm CO_{2}$
laser. The total laser power was controlled by passing the output
light through an acousto-optic modulator (AOM). The first order of
the modulator was then directed into the telescope to expand the
beam. The optical arrangement used in this experiment is as shown
in Fig.\thinspace \ref{exp_schematic}. For our atomic source we used
a magneto-optic trap (MOT), formed with a
20 G/cm magnetic field gradient, and by two 5 cm diameter, 20 mW
beams. Each beam made three passes through the chamber and was
detuned 15 MHz below the $F=2\: \rightarrow \: F'=3$ transition of
the D2 line of ${\rm Rb\  87}$. Repumping light tuned to the
$F=1\: \rightarrow \: F'=2$ transition propagated with one of the
trapping beams. We were able to capture about $2\times 10^{7}$ atoms
with this arrangement.
One of the most difficult aspects involved in the setup of a FORT
is the beam alignment. Since the FORT light is so far from
resonance, no fluorescence is induced and it is hard to determine
whether the FORT beam is overlapping with the MOT or not. To overcome this
difficulty we have devised a method which allows us to observe the
position of the $\rm CO_{2}$ laser beam in real time directly on
an inexpensive CCD camera that normally monitors the MOT. To
accomplish this it is necessary to improve the contrast between
the atoms trapped in the MOT and those trapped in the FORT.
Several techniques will work, for example, increasing the detuning
of the MOT light from resonance, or reducing the intensity of the
MOT light. A similar effect is obtained if these operations are
performed on the repumping light. With any one of these methods,
the brightness of the MOT and the effect of its near resonant
light on pushing atoms out of the FORT, can be lessened.
However, there can still be enough near-resonant light present in
the MOT beams to cause atoms that are contained in the FORT to
fluoresce and hence make the FORT beam visible. Using these
techniques greatly simplifies alignment of our $\rm CO_{2}$ laser
beam, turning a task which could previously take several days into
one that can be performed in minutes.

To load the FORT with atoms we apply the following procedure.
First the MOT is loaded for $30$ seconds from the background vapor
while at the same time the $\rm CO_2$ laser remains switched on.
Then, as a key step in efficiently loading the FORT, we reduce the
repump intensity by a factor of 50 compared to its initial value
to make a temporal dark SPOT \cite{Wieman_Phys_Rev_A,
ketterle_darkspot}. This strong reduction in the repump power
occurs $50$ to $70\thinspace \rm ms$ before switching off the MOT
trapping beams. Alongside reducing the repump power, we jump the
trapping beam detuning to $-80\thinspace \rm MHz$ for further
laser cooling and to counteract the detuning change induced by the
light shift of the ${\rm CO_2}$ beam. If we did not jump the detuning, atoms in
the region of the FORT would see the MOT beams positively detuned, thus
reducing the effectiveness of the MOT. Finally, after the main MOT
beams have been extinguished, we adiabatically switch off the MOT
magnetic field. The earliest time that we can image the FORT is
$100\thinspace \rm ms$ after releasing the MOT. This ensures that
any of the untrapped atoms have sufficient time to fall away from
the FORT under the influence of gravity. Both the MOT and the FORT
are destructively imaged by observing the absorption of a resonant
probe laser which passes through the atom cloud and is incident on
a CCD camera. By integrating the optical density across the atom
cloud we are able to determine the number of trapped atoms. When
imaging the FORT, the $\rm CO_2$ laser beams are switched off 3.5
ms before the image is taken to allow the cloud of atoms to expand
to a size which is significantly above the resolution of the
optical system.

\section{Results and Discussion}
\begin{figure}
\centerline{\psfig{figure=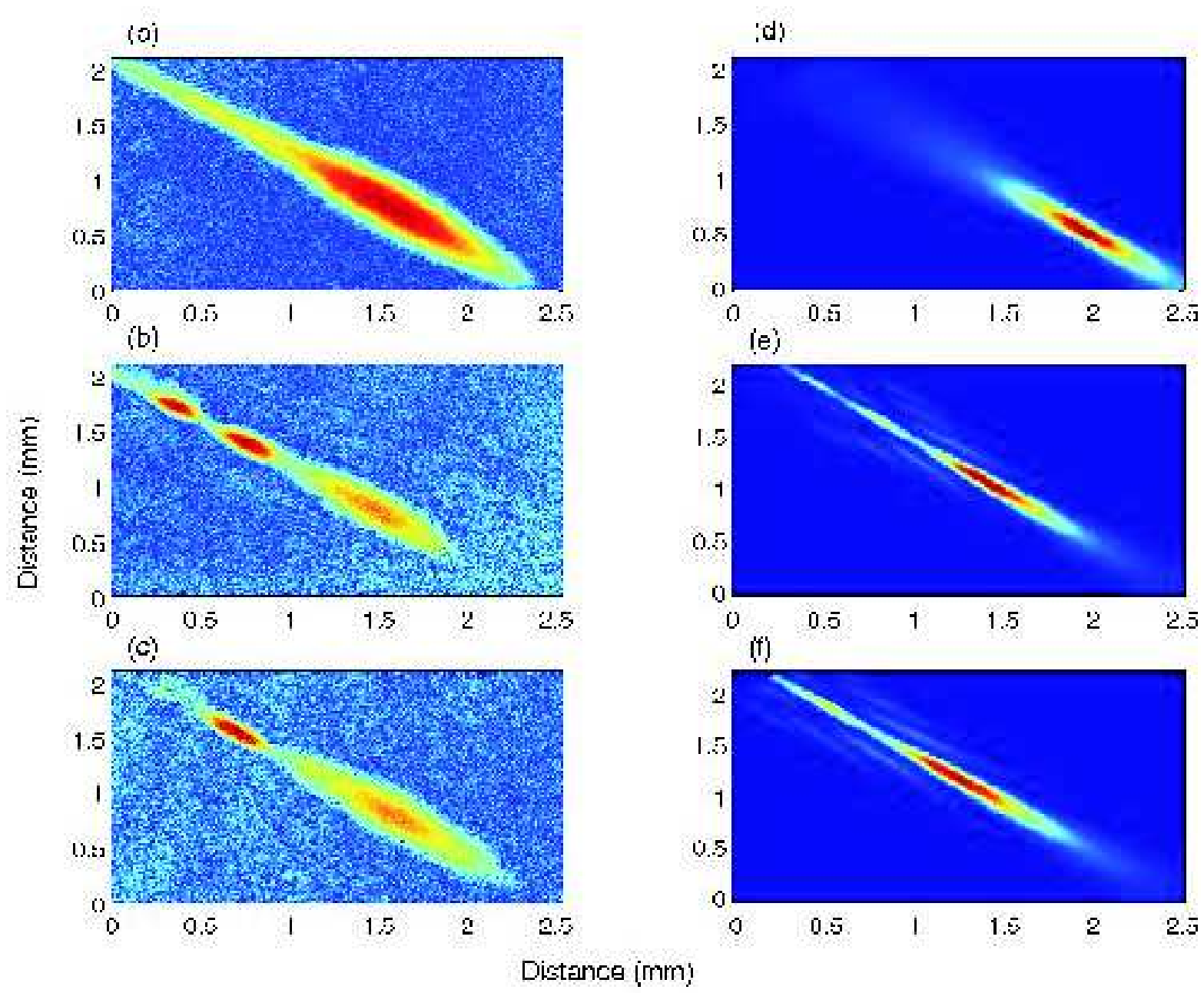}} \caption[]{\label{fort}
Absorption images of the FORT with their corresponding theoretical
images. The abscissa and ordinate are distances in millimeters in
each image. (a) The spherical aberration $\beta$ is less than a
wavelength because of the small size of the $\rm{CO_2}$ beam on
the primary lens. (b) The spherical aberration is around 12.6
wavelengths (c) The beam size on the primary lens is larger and
produces strong aberration of around 18.2 wavelengths. Figures
(d), (e) and (f) are the theoretical images corresponding to the
cases (a), (b) and (c) respectively. Note that atom clouds in the
experiment have expanded for 3.5 {\it ms} before the image was
taken}
\end{figure}

Using the procedure described in Section 3 we have been able to
load several micro-optical traps created by the aberration pattern
of the meniscus lens. Interestingly, our observations show that
there are approximately 10 sites capable of holding the atoms when
the beam diameter is similar to the size of the primary lens, in
very good agreement with our simulation results. For a given
amount of spherical aberration $\beta$, the separation between the
micro-traps decreases as one moves away from the focus and towards
the primary lens. Therefore the number of micro-traps being loaded
from a MOT at a given time depends where the MOT and FORT overlap
each other. Usually 3 micro-traps are loaded by placing the atomic
cloud of the the MOT close to the gaussian focus of the lens. This
can be increased to 4 or 5 micro-traps by moving the MOT a few
millimeters towards the lens (to move the MOT we change the
currents in the nulling coils designed to cancel out stray
magnetic fields). Fig.\thinspace \ref{fort} shows three absorption
images of the FORT and their corresponding simulated potentials at
two extreme separations of the telescope lenses and one
intermediate separation.
The lower panel of Fig.\thinspace \ref{fort} shows that the central FORT
and one micro-trap are loaded.
This happens when there is higher spherical aberration $\beta$.
Our simulations show that the separation between the peaks is greater when
there is a high spherical aberration. So in the lower panel of Fig.\thinspace \ref{fort}, the spacial extent of
the MOT is such that it could load only one micro-trap along with the central FORT. For the parameters of
Fig.\thinspace \ref{fort}(c), we found from equation \thinspace \ref{beta} that
the spherical aberration $\beta$ is around 18.2 wavelengths.
The central panel of Fig.\thinspace \ref{fort} shows the central FORT
and two micro-traps that are loaded
when $\beta$ is around 12.6 wavelengths.
In the upper picture, the spherical
aberration is diminished by the small beam size on the primary
lens so that only the highly populated central FORT remains. This
higher population is due to the fact that for such cases the beam
is not focused tightly so that the capture
volume of the FORT is increased.
In the absence of the spherical
aberration the central FORT usually contains $10^6$ trapped atoms.
In the presence of the spherical aberration, the other micro-traps
usually have $2 \times 10^5$ atoms at $70\mu \rm K$ temperature.

\begin{figure}
\begin{center} \mbox{\epsfxsize 4.0in\epsfbox{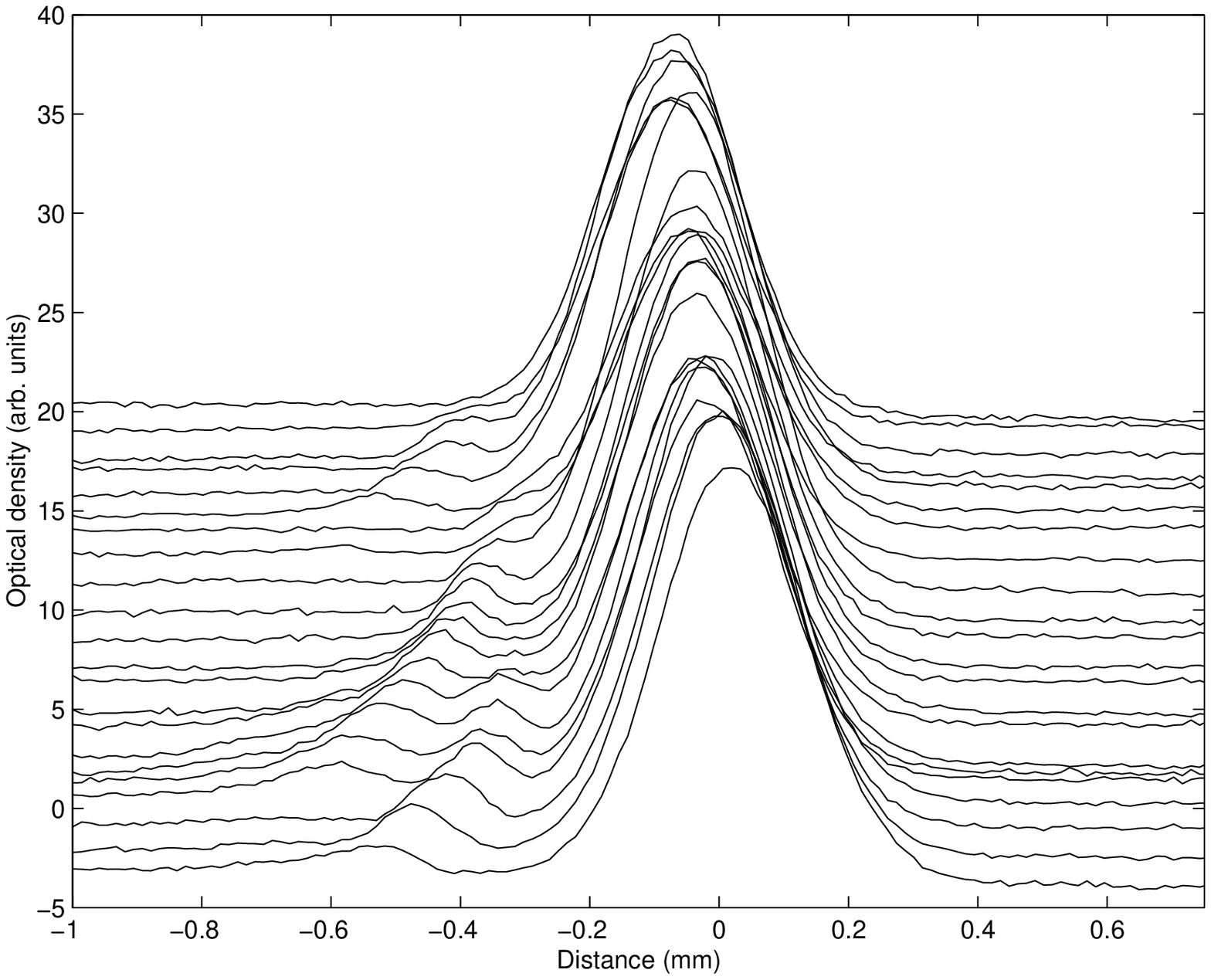}} \end{center}
\caption{Experimentally observed intensity integrated
perpendicular to the optical axis when the aberration is only due
to the telescope lenses. Each curve is for various telescope lens
separations.} \label{data}
\end{figure}

According to our numerical simulations, the spherical aberration
contributions from the telescope lenses can also alter the
intensity pattern close to the focal plane of the primary lens.
Our telescope lenses are not corrected for the spherical
aberration. To demonstrate this point the meniscus lens was
replaced with an aspheric lens corrected for spherical aberrations
so that the primary lens did not alter the wave front because of
the spherical aberration. Different combinations of lenses that
made up the telescope were tested, however all of them showed a
similar pattern. Thus, here we shall present only one set up in
which we used two plano convex lenses, both with 12.7cm focal
length. The separation of the telescope lenses was initially set equal to
24cm  and was then decreased in 3mm steps. Fig.\thinspace
\ref{data} shows the observed intensities along the optical axis
of the primary lens as the telescope separation was varied. An
offset has been added to each profile to improve the readability.
The sequence from the top is in order
of increasing distance between the telescope lenses. This figure
shows that a micro trap is created  from the central FORT and
starts to move away from it as we increase the separation of the
telescope lenses. This is because as the beam size on the second
lens increases so does the spherical aberration. From
Fig.\thinspace \ref{data} it can be noted that after the seventh step
of increment in the telescope separation a second micro-trap
emerges from the central FORT and moves away. This happens while
the first micro-trap has travelled far enough so that atoms are no
longer loaded into it. The second micro-trap moves away with
increasing telescope lens separation until the fifteenth step when
a third micro-trap emerges from the central trap and starts to
travel towards the second micro-trap. These two micro-traps
coexist for a few more increments in the separation until the
second micro-trap fails to load atoms. Since less atoms are loaded
into the micro-traps of Fig.\thinspace \ref{data} compared to Fig.
\thinspace \ref{fort}, we infer that the meniscus
lens produces more spherical aberration than the telescope lenses.

\section{Conclusion}
We have shown that the intensity distribution produced by a lens
that is not corrected for spherical aberration can be used to
prepare a potential to realize micro-optical traps. A beam from a
$\rm {CO_2}$ laser focused with such a lens was employed to load
$\rm {Rb\ 87}$ atoms into the micro-optical traps formed by the
intensity maxima of the spherical aberration pattern. Such high
density ($10^{13}$ atoms/$\rm{cm}^3$) samples of cold atoms are of
interest for a wide range of experimental studies including
evaporative cooling, cold collisions and quantum information
processing with ultra cold Rydberg atoms \cite{lucin}.
Furthermore, the variable  separation of the micro-traps could be
used to control the dipole-dipole interactions between the atoms
in adjacent traping sites. Also increasing the separation of the
micro-traps up to a few hundred microns makes the task of
individually addressing the different micro-traps relatively
straight forward with existing optical techniques. These
properties are of great interest for quantum information
processing proposals for neutral atoms. Another possible
experiment would be to construct an atom interferometer using the
micro-traps. One could take a BEC formed using evaporative cooling
in a single focused laser beam \cite{osu bec} and then by changing
the separation of the telescope lenses split off a sub group of
BEC atoms. By simply setting the telescope separation back to the
initial value the two BEC groups can be recombined making an
interference pattern which depends on the phase difference
accumulated between the wavefunctions. An analysis of such an
interference pattern releases information about the mechanisms
effecting the phase of the transported BEC. For example, if the
second $\rm{CO_2}$ beam propagates in the vertical direction a
phase will be induced to the wavefunction proportional to the
change in the gravitational potential of the moving group.
Therefore the final interference pattern contains information that
could be used to probe gravity.

\section{Acknowledgements}
We wish to acknowledge Brian Timmons for his contributions to the experimental
set up.


\Bibliography{50}
\bibitem{ChuFORT} C.S.\thinspace Adams, H.J.\thinspace Lee, N.\thinspace Davidson, M.\thinspace Kasevich,
and S.\thinspace Chu, Phys.\thinspace Rev.\thinspace
Lett.\thinspace {\bf 74}, 3577 (1995).

\bibitem{Hansch&Weitz_FORT} S.\thinspace Friebel, C.\thinspace D'Andrea, J.\thinspace Walz, M.\thinspace
Weitz, and T.W.\thinspace H\"{a}nsch, Phys.\thinspace
Rev.\thinspace A {\bf 57}, R20 (1998).

\bibitem{ChapmanBEC} M.D.\thinspace Barrett, J.A.\thinspace Sauer,
and M.S.\thinspace Chapman, Phys.\thinspace Rev.\thinspace
Lett.\thinspace {\bf 87}, 010404 (2001).

\bibitem{CsBEC} T.\thinspace Weber J.\thinspace Herbig, M.\thinspace Mark, H.C.\thinspace
N\"agerl and R.\thinspace Grimm, Science {\bf 299}, 232 (2003).

\bibitem{YbBEC} Y.\thinspace Takasu, K.\thinspace Maki, K.\thinspace Komori, T.\thinspace Takano,
K.\thinspace Honda, M.\thinspace Kumakura, T.\thinspace Yabuzaki, and Y.\thinspace Takahashi,
Phys.\thinspace Rev.\thinspace Lett.\thinspace {\bf 91}, 040404 (2003).

\bibitem{Thomas_FermiGas} S.R.\thinspace Granade, M.E.\thinspace Gehm, K.M.\thinspace O'Hara,
and J.E.\thinspace Thomas, Phys.\thinspace Rev.\thinspace
Lett.\thinspace {\bf 88}, 120405 (2002).

\bibitem{WeitzPRL_BEC} G.\thinspace Cennini, G.\thinspace Ritt, C.\thinspace Geckeler, and M.\thinspace
Weitz, Phys.\thinspace Rev.\thinspace Lett.\thinspace {\bf 91}
240408 (2003).

\bibitem{Brennen} G.K.\thinspace Brennen, C.M.\thinspace
Caves, P.S.\thinspace Jessen and I.H.\thinspace Deutsch,
Phys.\thinspace Rev.\thinspace Lett.\thinspace {\bf 82}, 1060 (1999).

\bibitem{Borion}D.\thinspace Boiron, A.\thinspace Michaud,
J.M.\thinspace Fournier, L.\thinspace Simard, M.\thinspace
Sprenger, G.\thinspace Grynberg, and C.\thinspace Salomon,
Phys.\thinspace Rev.\thinspace A. {\bf 57}, R4106 (1998);
R.\thinspace Newell, J.\thinspace Sebby, and
T.G.\thinspace Walker, Opt.\thinspace Lett.\thinspace {\bf 28}, 14, (2003).

\bibitem{Dumke} R.\thinspace Dumke, M.\thinspace Volk,
T.\thinspace M\"{u}ther, F.B.J\thinspace Buchkremer, G.\thinspace
Birkl, and W.\thinspace Ertmer, Phys.\thinspace Rev.\thinspace
Lett.\thinspace {\bf 89}, 097903 (2002).

\bibitem{Birkl}G.\thinspace Birkl, F.B.J\thinspace Buchkremer, R.\thinspace Dumke,
and W.\thinspace Ertmer,\thinspace Opt.\thinspace Commun.\thinspace {\bf
191}, 67 (2001).

\bibitem{Buchkremer} F.B.J\thinspace Buchkremer,\thinspace {\it et al.}, Laser
\thinspace Phys.\thinspace {\bf 12}, 736 (2002).

\bibitem{peil} S.\thinspace Peil, J.V.\thinspace Porto, B.L.\thinspace
Tolra, J.M.\thinspace Obrecht, B.E.\thinspace King, M.\thinspace
Subbotin, S.L.\thinspace Rolston and W.D.\thinspace Phillips,
Phys.\thinspace Rev.\thinspace A, {\bf 67}, 051603, (2003).

\bibitem{Evance} L.R.\thinspace Evans and C.G.\thinspace Morgan,
Nature, {\bf 219}, 712, (1968).

\bibitem{Evance-prl} L.R.\thinspace Evans and C.G.\thinspace Morgan,
Phys.\thinspace Rev.\thinspace Lett.\thinspace {\bf 22}, 1099, (1969).

\bibitem{smith} L.M.\thinspace Smith, J.\thinspace Opt.\thinspace Soc.\thinspace
Am.\thinspace A., {\bf 6}, 1049 (1989)

\bibitem{Nagerl} H.C.\thinspace N\"{a}gerl, D.\thinspace
Leibfried, H.\thinspace Rohde, G.\thinspace Thalhammer,
J.\thinspace Eschner, F.\thinspace Schmidt-Kaler, and R.\thinspace
Blatt, Phys.\thinspace Rev.\thinspace A, {\bf 60}, 145 (1999).

\bibitem{born} M.\thinspace Born and E.\thinspace Wolf, "Principles of optics",
7 ed., p.519, 'Cambridge University Press' (1999).

\bibitem{yashida} A.\thinspace Yoshida and T.\thinspace Asakura, Opt.\thinspace
Comm.\thinspace {\bf 25}, 133 (1978).

\bibitem{welfold} W.T.\thinspace Welfold, ''Aberrations of the
symmetrical optical system'', p.192, 'Academic Press', London
(1974).

\bibitem{Wieman_Phys_Rev_A} S.J.M.\thinspace Kuppens, K.L.\thinspace Corwin, K.W.\thinspace
Miller, T.E.\thinspace Chupp and C.E.\thinspace Wieman,
Phys.\thinspace Rev.\thinspace A, {\bf 62}, 013406 (2000).

\bibitem{ketterle_darkspot} W.\thinspace Ketterle, K.B.\thinspace Davis, M.A.\thinspace Joffe, A.\thinspace Martin and
D.E.\thinspace Pritchard, Phys.\thinspace Rev.\thinspace Lett.\thinspace {\bf 70}, 2253 (1993).

\bibitem{lucin} M.D.\thinspace Lukin, M.\thinspace Fleischhauer,
R.\thinspace Cote, L.M.\thinspace Duan, D.\thinspace Jaksch, J.I. \thinspace
Cirac and P.\thinspace Zoller, Phys.\thinspace Rev.\thinspace Lett. {\bf 87},037901 (2001).

\bibitem{osu bec} We have recently realized an all optical BEC and successfully
transferred it into the focus of a CO$_2$ laser beam. The details
to be published elsewhere.

\endbib
\end{document}